# Two-step excitation of fluorescent proteins with real intermediary states

**J. David Wong-Campos,[1,*] Dalia P. Ornelas-Huerta,[1] and Mackenzie Dion[1]**

[1] *Triplet Imaging, 325 Vassar St., Cambridge, Massachusetts, 02139*
*\*jwongcampos@gmail.com*

**We demonstrate sequential two-photon fluorescence microscopy using forbidden state transitions. Nonlinear red excitation leads to green fluorescence in live cells expressing eYFP, maintaining optical sectioning and allowing deep tissue imaging with simple optical systems.**

Multiphoton microscopy has revolutionized biological imaging by enabling deep tissue penetration and optical sectioning [**1**]. However, its widespread adoption remains limited by the requirement for complex and expensive ultrafast laser systems. Here, we demonstrate a novel approach to nonlinear excitation that harnesses real intermediate states rather than virtual transitions, enabling multiphoton imaging with both pulsed and continuous-wave sources. This method, which we term two-photon prime (2p-prime), exploits the dynamics of dark states [**2**] to achieve efficient nonlinear excitation at unconventional wavelengths. Following measurements of triplet states in fluorescent proteins [**3**], we apply this new approach to image HEK 293T cells expressing cytosolic enhanced Yellow Fluorescent Protein (eYFP).

    The home-built microscope [Fig. 1(a)] uses a supercontinuum laser (SC, FIU-15, NKT) providing picosecond pulses at variable repetition rates and wavelength, or a continuous wave laser at 660 nm (LSR660SMFC-80, Civil laser). When using the supercontinuum laser, light is filtered by a long-pass (F1: FGL630M, short-pass F2: FESH1000, Thorlabs) and a tunable filter pair (TFP: LF104550/LF104555, Delta). A scanning unit consisting of a galvo-galvo pair (SU, QS7XY-AG, Thorlabs) with scan lens (SL, SL50-CLS2, Thorlabs) and tube lens (TL, TTL200-B, Thorlabs) delivers the beam to the objective (UPLXAPO 40X, 0.95 NA, Olympus). A dichroic mirror (D1, 610lpxr-t3-Di, Chroma) directs fluorescence through an emission filter (F1, ET535/70m, Chroma) to a silicon photomultiplier

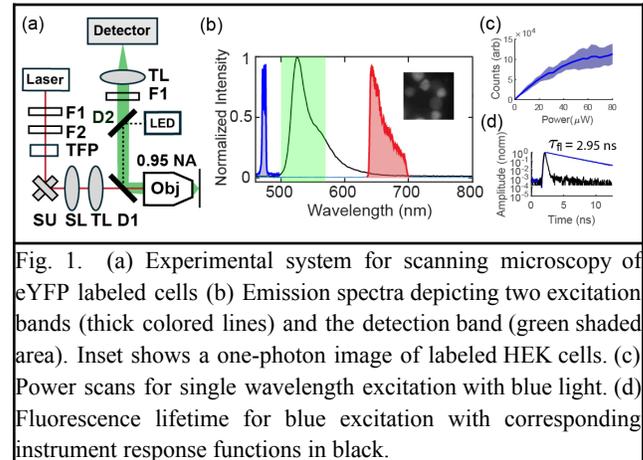

Fig. 1. (a) Experimental system for scanning microscopy of eYFP labeled cells (b) Emission spectra depicting two excitation bands (thick colored lines) and the detection band (green shaded area). Inset shows a one-photon image of labeled HEK cells. (c) Power scans for single wavelength excitation with blue light. (d) Fluorescence lifetime for blue excitation with corresponding instrument response functions in black.

(Detector, PDA45, Thorlabs) via tube lens (TL, AC254-100-A, Thorlabs). Data acquisition (NI USB-6363, National Instruments) and control software (ScanImage, MBF Bioscience) process the signals. Z-stacks use a piezo scanner (PFM450E, Thorlabs). For lifetime measurements, we use a photon counter (FastGatedSPAD, MPD) and time tagger (Time Tagger Ultra, Swabian). One-photon (1P) imaging uses an LED (LED, M470L5, Thorlabs), a dichroic filter (D2, FF495-Di, AVR), and camera (BFS-U3-200S6M-C, FLIR).

    Human embryo kidney (HEK) cells were transfected with eYFP plasmid (Twist biosciences) using Mirus Bio reagent. eYFP was chosen due to the reported presence of long lived states [**4**]. Twenty-four hours post-transfection, cells were trypsinized, resuspended in DMEM, and plated in Poly-D-Lysine-coated glass bottom dishes (0.085-0.115 mm thickness) for imaging.

    Figure 1(b) shows the fluorescence emission and excitation bands. Blue light (475 nm) excitation exhibited typical one-photon saturation at high powers [Fig. 1(c)] with a fluorescence lifetime of 2.95 ns [Fig. 1(d)]. As shown in Fig. 2(a-b), excitation with a focused 665 nm beam revealed a nonlinear power dependence with an

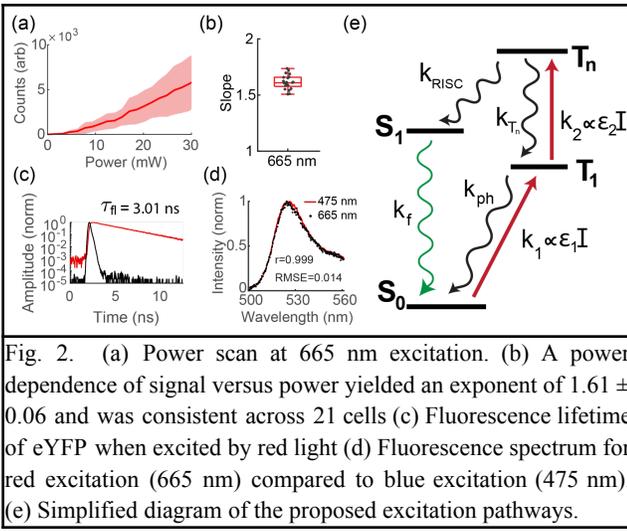

Fig. 2. (a) Power scan at 665 nm excitation. (b) A power dependence of signal versus power yielded an exponent of 1.61 ± 0.06 and was consistent across 21 cells (c) Fluorescence lifetime of eYFP when excited by red light (d) Fluorescence spectrum for red excitation (665 nm) compared to blue excitation (475 nm). (e) Simplified diagram of the proposed excitation pathways.

exponent of 1.61 ± 0.06, as measured across several cells (n = 21 cells). To validate the signal's origin and rule out possible contaminants, we compared the fluorescence lifetime and emission spectra between red and 1P excitation. We observed excellent agreement between both excitation schemes [Fig. 2(c-d)], confirming that red excitation generated fluorescence from the same excited state and protonation species as blue light.

The non-integer slope indicated excitation through real intermediate states, in contrast to the quadratic dependence characteristic of two-photon excitation through virtual states in eYFP [5]. A simplified four-level model [Fig. 2(e)] consisting of excitation rates proportional to the intensity $k_{1,2}=\varepsilon_{1,2}I$, with extinction coefficients $\varepsilon_{1,2}$ of the intermediary real states $T_{1,n}$, decays from the states $T_1$ and $T_n$ of $k_{ph}$ and $k_{T_n}$ of the same manifold, a rate of intersystem crossing between $T_n$ and $S_1$, $k_{RISC}$, and $k_f$ for the fluorescence rate explains this behavior: a sequential multi-step process is facilitated by long-lived intermediate states. Assuming faster decay within the same manifold versus intersystem crossing ($k_{T_n} \gg k_{RISC}$) the solution to the steady state equations yields:

$$Power\ exponent = 2 - \frac{1}{1+k_{ph}/k_2},$$

which shows how the power dependence is modified by competition between $T_1$ decay ($k_{ph}$) and excitation rate to $T_n$ ($k_2$).

We demonstrate the imaging capabilities of this 2p-prime scheme in Fig. 3(a). In three-dimensional cell aggregates, we compared one-photon widefield microscopy with picosecond and CW 2p-prime excitation at 665 nm [Fig. 3(a)]. With average powers of 18 mW (ps) and 42 mW (CW) at the sample and dwell times of 5 μs and 20 μs, respectively, we achieved optical sectioning at depths not attainable by widefield one-photon excitation and lower power levels than previously reported CW-2P [6]. Line profiles through a single cell revealed an improved contrast at depth [Fig. 3(b)], while maintaining reasonable photobleaching characteristics [Fig. 3(c)] compared to traditional 2P [7]. Notably, CW excitation required only 10 times higher average power than pulsed excitation to achieve comparable image quality [Fig. 3(d)].

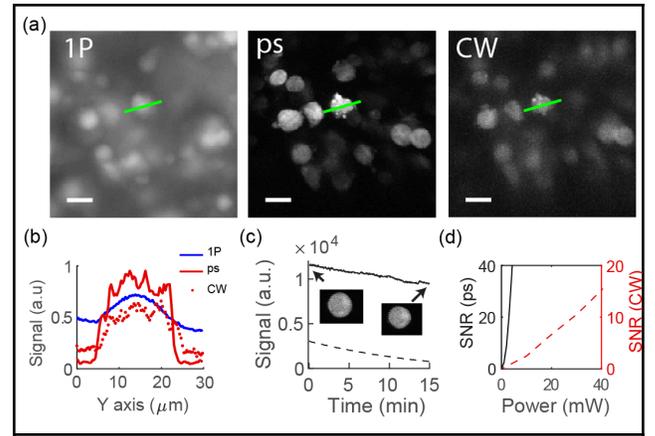

Fig. 3. (a) Comparison between widefiled 1P illumination, 2p-prime with picosecond pulses (ps), and continuous wave (CW) lasers at 0.4, 18 and 42 mW average power at the sample. Scanned images for 2p-prime were taken at 5 μs (ps) and 20 μs (CW) dwell times in a 512×512 pixel area. (b) Line profile plot of the green line in (a) showing improved contrast at depth. (c) Photobleaching at 665 nm excitation, dashed line is the background bleaching. (d) Signal-to-Noise vs. power comparison between ps and CW at 20μs dwell time. Scale bar is 20 μm.

The demonstration of nonlinear excitation through real intermediate states opens several promising avenues for both fundamental research and applications. We hypothesize our red light pumping out of the triplet state scheme may also be protecting against bleaching as previously reported by IR light on several green fluorescent proteins [8]. Our approach provides high SNR at average powers similar to traditional 2P schemes with simpler systems and may allow novel microscopy geometries where a highly efficient nonlinear process could take place across larger areas, enabling light-sheet applications of large volumes without mechanical scanning. Furthermore, targeted optimization of fluorescent proteins for this mechanism provides an alternative venue to reduce excitation powers, potentially allowing wider accessibility.

**Funding.** This research was funded by Triplet Imaging, Inc.

**Acknowledgment.** We thank He Tian for cell culture help and Isaac Garcia Reyes for useful discussions.

**Disclosures.** J.D.W-C. and M.D are cofounders of Triplet Imaging, Inc.

**Data Availability.** Data may be obtained from the authors upon reasonable request